\documentclass[prb,twocolumn,amssymb,amsmath,nobibnotes]{revtex4}

\usepackage{graphicx,color}
\usepackage{subfigure}
\usepackage{dcolumn}
\usepackage{bm}
\usepackage{threeparttable}

\begin{document}

\newenvironment{figurehere}
  {\def\@captype{figure}}
  {}

\title{Dielectric properties and lattice dynamics of $\alpha$-PbO$_{2}$-type TiO$_{2}$: The role of soft phonon modes in pressure-induced phase transition to baddeleyite-type TiO$_{2}$}
\author{Yongqing Cai}
\author{Chun Zhang}
\author{Yuan Ping Feng}

\email{phyfyp@nus.edu.sg}
\affiliation{
	Department of Physics, National University of Singapore, 2 Science Drive 3, Singapore 117542\\}
\date{\today}

\begin{abstract}
   Dielectric tensor and lattice dynamics of $\alpha$-PbO$_{2}$-type TiO$_{2}$ have been investigated using the density functional perturbation theory, with a focus on responses of the vibrational frequencies to pressure. The calculated Raman spectra under different pressures are in good agreement with available experimental results and the symmetry assignments of the Raman peaks of $\alpha$-PbO$_{2}$-type TiO$_{2}$ are given for the first time. In addition, we identified two anomalously IR-active soft phonon modes, $B_{1u}$ and $B_{3u}$, respectively, around 200 cm$^{-1}$ which have not been observed in high pressure experiments. Comparison of the phonon dispersions at 0 and 10 GPa reveals that softening of phonon modes also occurs for the zone-boundary modes. The $B_{1u}$ and $B_{3u}$ modes play an important role in transformation from the $\alpha$-PbO$_{2}$-type phase to baddeleyite phase. The significant relaxations of the oxygen atoms from the Ti$_{4}$ plane in the Ti$_{2}$O$_{2}$Ti$_{2}$ complex of the baddeleyite phase are directly correlated to the oxygen displacements along the directions given by the eigenvectors of the soft $B_{1u}$ and $B_{3u}$ modes in the $\alpha$-PbO$_{2}$-type phase.

\end{abstract}

\maketitle

\section{INTRODUCTION}
    Titanium dioxide (TiO$_2$) has attracted much interest for a long time due to its potential applications in photocatalysis. Besides the three well studied ambient phases (rutile, anatase, and brookite),\cite{Amorphization_PRB1,Amorphization_PRB2,Exp_Cal_Anatase} interest in TiO$_2$ has also extended to high pressure phases to fully explore its novel structural and electronic properties.\cite{B0,Khatatbeh,Anatase_Size2} For instance, application of high pressure on TiO$_{2}$ at room temperature induces a series of high-pressure phases with hardness approaching that of diamond.\cite{Ultrastiff} The cotunnite-type TiO$_{2}$ synthesized at pressure ($P$) above 60 GPa and temperature above 1000 K is believed to be the hardest oxide yet discovered.\cite{cotunnite} In addition, density functional theory calculation predicted that the high-pressure modification of TiO$_{2}$ could lead to a polymorph which responds to visible light,\cite{Visible} extending the photocatalyst property of TiO$_2$ from ultraviolet to visible light.

   The high pressure induced structural transition of TiO$_{2}$ generally follows these sequences: rutile or anatase $\rightarrow$ $\alpha$-PbO$_{2}$-type (space group: $Pbcn$) $\rightarrow$ baddeleyite type (space group: $P2_{1}/c$) $\rightarrow$ orthorhombic structure (OI, space group: $Pbca$) $\rightarrow$ cotunnite type (OII, space group: $Pnma$).\cite{OI,Anatase_70} Critical pressures of transitions between two successive phases are found to be strongly dependent on the size of the sample and the experimental temperature.\cite{Amorphization_unusual,Anatase_Size1} Generally, at room temperature, the critical pressure of the transition from rutile (anatase) to $\alpha$-PbO$_{2}$-type phase is located in the region between 2.6 and 10 GPa (2.5 and 7 GPa).\cite{Boundary,alpha_PbO2_TS1,Anatase_Size2,Anatase_TS,Anatase_Size1} Interestingly, it was found that all the high pressure phases recover to the $\alpha$-PbO$_{2}$-type phase rather than the initial rutile or anatase phase upon decompression to ambient conditions.\cite{TiO2_23}

   Since the $\alpha$-PbO$_{2}$-type phase plays a bridging role in connecting the ambient TiO$_{2}$ phase and the higher pressure phases, a number of studies have been carried out on phase transformations involving the $\alpha$-PbO$_{2}$-type phase, such as rutile(anatase) $\rightleftharpoons$ $\alpha$-PbO$_{2}$\cite{Boundary,Anatase_Shock} and $\alpha$-PbO$_{2}$ $\rightleftharpoons$ baddeleyite.\cite{TiO2_23,alpha_PbO2_TS2,baddeleyite_laser} Among these, the mechanism for transition from the rutile phase to the $\alpha$-PbO$_{2}$-type phase has been well understood, and it can be described by the 1/2[01$\overline{1}$] shear of every other hexagonal close-packed (hcp) plane (011) of the rutile phase.\cite{alpha_PbO2_TS1} A soft phonon mode ($B_{1g}$) with a negative pressure derivative was found in the vibrational frequencies\cite{Strained_rutile} which is crucial to drive the initial structure to a CaCl$_{2}$-type intermediate structure during this transition.\cite{Pathways,Rutile_Cacl2} A similar mechanism was proposed by Sekiya for the anatase to $\alpha$-PbO$_{2}$ phase transition.\cite{Anatase_TS} In this case, a soft mode($E_{g}$) observed around 200 cm$^{-1}$ in the Raman spectra was suggested to be closely related to the phase transition. However, for the subsequent $\alpha$-PbO$_{2}$ $\rightleftharpoons$ baddeleyite phase transition, the understanding of which should form the basis of investigating the transitions of TiO$_{2}$ to the higher pressure phases (OI and OII), no phonon softening was observed, and the mechanism of this displacive phase transition has not been reported.

   In this study, we carry out density functional perturbation theory (DFPT) calculations to investigate the dielectric property and Born effective charge tensor of $\alpha$-PbO$_{2}$-type TiO$_{2}$. In addition, we study its lattice dynamical properties and their variations with pressure to identify soft phonon modes associated with pressure induced structural instability. The Raman intensity is calculated and presented for pressure from 0 to 10 GPa and compared with the experimental spectrum. The symmetry assignments of the Raman peaks for $\alpha$-PbO$_{2}$-type TiO$_{2}$ are given for the first time. In contrast to the Raman active soft modes of rutile and anatase phases, two IR active soft phonon modes ($B_{1u}$ and $B_{3u}$) with frequencies around 200 cm$^{-1}$ are identified for the $\alpha$-PbO$_{2}$-type phase. Based on the analysis of the two soft modes, we conclude that the Ti$_{3}$O triangular complex and the Ti$_{2}$O$_{2}$Ti$_{2}$ complex are the relevant structural units for understanding the phase transformation of $\alpha$-PbO$_{2}$-type TiO$_{2}$ $\rightleftharpoons$ baddeleyite TiO$_{2}$.

\section{METHODS}
   All calculations were carried out using the CASTEP code \cite{CASTEP_1,CASTEP_2} with the Ceperley-Alder local density functional approximation.\cite{CA_LDA} Brillouin zone integration was performed using the Monkhorst-Pack $\emph{k}$-point sampling\cite{MK} with $4\times 4\times 6$ and $3\times 3 \times 3$ grids for $\alpha$-PbO$_{2}$-type and baddeleyite phases, respectively. Norm-conserving pseudopotentials\cite{Norm-conserving} were generated by the Opium package\cite{OPIUM} and employed throughout the study. For Ti atom, the pseudopotential was constructed to include the $3s$ and $3p$ orbitals as semi-core states. The core radii $r_c$ for the pseudo wavefunctions were tuned to be 1.35 ($3s$), 1.35 ($3p$), 1.45 ($3d$), and 1.35 ($4s$) a.u. to ensure a good transferability of the potentials. For O atom, the $r_c$ values are 1.4 a.u. for all the $2s$, $2p$ and $3d$ valence states. The structures were relaxed until the forces exerted on the atoms are less than 0.01 eV/\AA, and the stress exerted on the cell is less than 0.02 GPa.

   The vibrational frequencies at the $\Gamma$ point, the transverse-optical/longitudinal-optical (TO/LO) splitting of the zone center optical modes, Raman spectra, and dielectric properties were calculated using the density functional perturbation theory.\cite{DFPT} Before calculations were carried out for the $\alpha$-PbO$_{2}$ and baddeleyite phases, phonon spectrum of rutile TiO$_{2}$ was calculated and compared with available results\cite{Strained_rutile}. It was found that for the chosen pseudopotentials, a plane-wave cutoff of 980 eV was sufficient to obtain reliable frequencies of the $\Gamma$ modes and this cutoff was used in subsequent calculations.

   To obtain the phonon dispersion curves of $\alpha$-PbO$_{2}$-type TiO$_{2}$, we first calculated the dynamical matrix based on a $4\times 4\times 4$ grid of $q$ points, with a (1/4,1/4,1/4) shift to include the $\Gamma$ point, using the density functional perturbation theory. The Fourier interpolation scheme was used to generate the dynamic matrix at an arbitrary $q$ point based on the aperiodic force constant matrix in real space produced by the cumulant sum method.\cite{Cumulant_sum} The long-range dipole-dipole interaction was also taken into account by using the calculated anisotropic Born effective charge tensor and dielectric tensor.

\section{RESULTS}
\subsection{Structure of $\alpha$-PbO$_{2}$-type TiO$_{2}$}

\begin{figure}
\includegraphics[width=8.0cm]{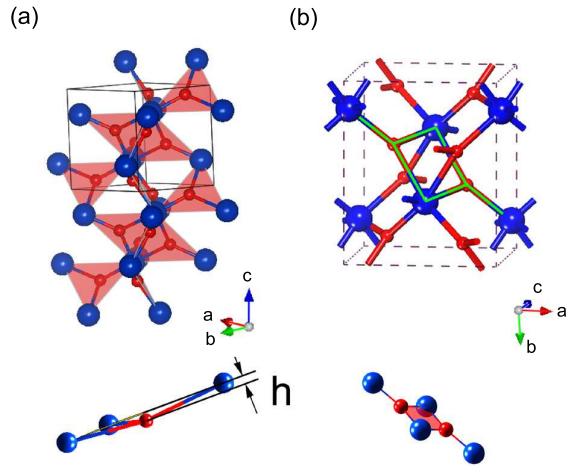}\\
\caption{(Color online) Crystal structure of $\alpha$-PbO$_{2}$-type TiO$_{2}$ viewed from different directions to show (a) the Ti$_{3}$O (filled triangles) and (b) Ti$_2$O$_2$Ti$_2$ complexes (green lines). $h$ denotes the distance of the O atom from the plane defined by the Ti atoms. Red/small balls and blue/big balls denote oxygen and titanium atoms, respectively.}
\label{Fig.1}
\end{figure}

      Since both the vibrational properties and phase transition of a compound are closely related to its atomic structure, we present first calculated structural information of the $\alpha$-PbO$_{2}$-type phase. The relaxed lattice parameters and atomic positions at zero pressure are given in Table I and are compared with available experimental values. The unit cell of $\alpha$-PbO$_{2}$-type TiO$_{2}$ contains four formula units with four equivalent Ti atoms and eight equivalent O atoms. As shown in Fig. 1, each Ti atom is surrounded by 6 adjacent oxygen atoms with a coordination number (CN) of 6. The O atom is bonded with three Ti atoms forming an approximately ``Y''-shaped Ti$_{3}$O triangular complex, and two Ti$_{3}$O complexes further form a Ti$_{2}$O$_{2}$Ti$_{2}$ complex (Fig. 1b) through sharing a Ti$_{2}$ edge, similar to the case of rutile TiO$_2$. There are four Ti$_{2}$O$_{2}$Ti$_{2}$ complexes in each unit cell. The centers of the four Ti$_{2}$O$_{2}$Ti$_{2}$ complexes are located at a corner (0,0,0), the center (1/2,1/2,1/2), an edge (0,0,1/2), and a face center (1/2,1/2,0) of the cubic unit cell, respectively, which are the four inversion centers of the unit cell. In contrast to rutile phase where Ti and O atoms forming the Ti$_{2}$O$_{2}$Ti$_{2}$ complex are all in the same plane, the Ti$_{3}$O and Ti$_{2}$O$_{2}$Ti$_{2}$ complexes in the $\alpha$-PbO$_{2}$ phase adopt a nonplanar structure. The two O atoms in the Ti$_{2}$O$_{2}$Ti$_{2}$ complex appear on opposite sides of the plane defined by the four Ti atoms (Ti$_{4}$ plane). The formation of these nonplaner Ti$_{3}$O complexes can be understood from the atomic relaxation driven by the $B_{1g}$ soft phonon mode in the rutile phase under pressure. \cite{Pathways,Rutile_Cacl2} These Ti$_{3}$O and Ti$_{2}$O$_{2}$Ti$_{2}$ complexes in the $\alpha$-PbO$_{2}$ phase play important roles in the analysis of Born effective charges and the description of $\alpha$-PbO$_{2}$ $\rightleftharpoons$ baddeleyite structural transformation which will be discussed in the following sections.

      The pressure induced structural modifications, including lattice parameters and atomic positions are calculated and results for $P=10$ GPa are presented in Table~I. The distance($h$) of O atom from the Ti$_{3}$ plane increases from 0.170 {\AA} at 0 GPa to 0.195 {\AA} at 10 GPa. The variations of $V/V_{0}$, where $V$ is the volume at a given pressure and $V_0$ is the volume at $P=0$ GPa, and the average Ti-O bond length with pressure are shown in Fig. 2. A linear decrease are seen for both quantities. The bulk modulus ($B_{0}$) and its pressure derivative $B^{'}$ are determined through fitting of $V/V_{0}$ to the  Birch-Murnaghan equation:\cite{Birch-Murnaghan}
\begin{eqnarray*}
P&=&\frac{3B_{0}}{2}\left[\left(\frac{V}{V_{0}}\right)^{-7/3}-\left(\frac{V}{V_{0}}\right)^{-5/3}\right]\times \\
& & \left\{1-\frac{3(4-B')}{4}\left[\left(\frac{V}{V_{0}}\right)^{-2/3}-1\right]\right\}
\end{eqnarray*}
      The fitted values of $B_{0}$ and $B^{'}$ are 248.6 GPa and 7.9, respectively. The bulk modulus obtained here is in good agreement with the experimental value of $258 \pm 8$ GPa.\cite{Laseralpha_PbO2}

\begin{table}[htbp]
\caption{Calculated structural parameters, atomic fractional positions, and densities of $\alpha$-PbO$_{2}$-type TiO$_{2}$ at 0 and 10 GPa.}
\begin{ruledtabular}
\begin{tabular*}{0.5\textwidth}{@{\extracolsep{\fill}}lcc}
         &     0 GPa                      &      10 GPa   \\   %
\hline
[a,b,c]  &   [4.531, 5.465, 4.870]        &   [4.487, 5.374, 4.804]            \\
         &   [4.531, 5.501, 4.906]\cite{Grey}  &                                    \\
         &   [4.61,  5.43,  4.87]\cite{Laseralpha_PbO2}   &                                    \\
Ti       &   (0, 0.176, 0.25)             &   (0, 0.179, 0.25)                 \\
         &   (0, 0.170, 0.25)\cite{Grey}       &                                    \\
O        &   (0.272, 0.384, 0.419)        &   (0.270, 0.389, 0.422)            \\
         &   (0.272, 0.381, 0.414)\cite{Grey}  &                                    \\
Density  &           4.401                &           4.581                    \\
         &           4.336\cite{Simons}            &
\end{tabular*}
\end{ruledtabular}
\end{table}

\begin{figure}
\includegraphics[width=8.0cm]{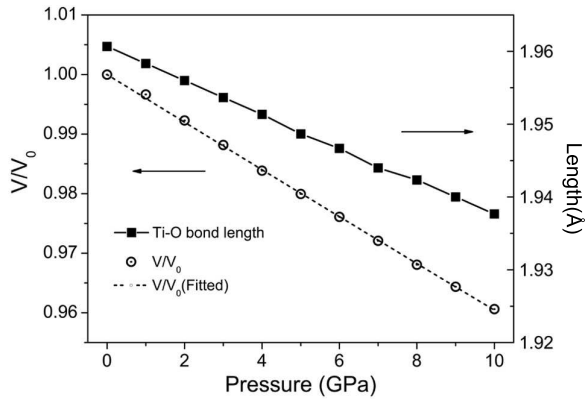}\\
\caption{Variations of $V/V_{0}$ and the average Ti-O bond length of $\alpha$-PbO$_{2}$-type TiO$_{2}$ with pressure. Here $V$ is the volume at pressure $P$, and $V_{0}$ is the volume at zero pressure.}
\label{Fig.2}
\end{figure}

\subsection{Born effective charges and dielectric tensors of $\alpha$-PbO$_{2}$-type TiO$_{2}$}
      Born effective charge (BEC) tensor quantifies macroscopic electric response of a crystal to the internal displacements of its atoms.\cite{Cai} The calculated Born effective charges of the $\alpha$-PbO$_{2}$-type TiO$_{2}$ are given in Table~II. It is known that BEC can be strongly influenced by the local environment of each atom, and the form and the number of independent elements in the BEC tensors are determined by the local atomic symmetry. Since $\alpha$-PbO$_{2}$-type possesses lower symmetry compared to rutile and anatase phases, the number of independent components in BEC tensors of Ti and O atoms in $\alpha$-PbO$_{2}$-type TiO$_{2}$ is much larger than those of rutile and anatase phases.\cite{BEC,anatase_phonon} In addition, in $\alpha$-PbO$_{2}$-type phase, since the Ti atoms occupy positions of higher local symmetry ($C_{2}$) than O atoms ($C_{1}$), the number of independent elements in the BEC tensor for Ti atom is smaller than that of O atom.

      The principal directions $(P_{1},P_{2},P_{3})$ of the Born effective charge tensor of the O atom are shown in Fig.~3, and the three principal values $(Z_{1}, Z_{2}, Z_{3})$ are $(-5.35, -1.19, -3.01)$. The component in $P_{1}$ direction is much smaller than the nominal ionic charge of the O ion $(Z=-2)$. The Born effective charge tensor of the O atom is strongly anisotropic with the absolute values of $Z_{1}$ and $Z_{3}$ much larger than that of $Z_{2}$. This behavior arises from the bonding structure of O atom in the Ti$_{3}$O complex. Each O atom in the Ti$_{3}$O complex is bonded to three adjacent Ti atoms by forming $sp^{2}$-like bonds. Therefore, the polarizations along the $P_{1}$ and $P_{3}$ directions are more sensitive to the displacement of O in the Ti$_{3}$ plane, compared to that in the normal direction ($P_{2}$). Since the Ti$_{3}$O complex in the $\alpha$-PbO$_{2}$ phase has a similar structure as the Ti$_{3}$O unit in the rutile case, the principal values of Born effective charge tensor for the O atom are comparable to the values $(-5.06, -1.34, -3.96)$ of the O atom in rutile phase.\cite{BEC}

\begin{table}[htbp]
\caption{Born effective charge, electronic ($\varepsilon_{\infty}$) and static ($\varepsilon_{0}$) dielectric tensors of $\alpha$-PbO$_{2}$-type TiO$_{2}$.}
\begin{ruledtabular}
\begin{tabular}{lc|lc}
\multicolumn{2}{c|}{Born effective charge tensors}  & \multicolumn{2}{c}{Dielectric tensors}         \\
                      \hline
Ti: &     $ \left(\begin{array}{ccc}
                               6.89  &  0   &  0.14    \\
                                   0 & 5.54 &    0     \\
                                0.36 &  0   &  6.68    \\
                            \end{array}
                      \right)$
                          & $\varepsilon_{\infty}:$  &
                      $\left(\begin{array}{ccc}
                                8.54  &  0   &   0     \\
                                0     & 6.95 &   0     \\
                                0     &  0   &  8.06   \\
                           \end{array}
                      \right)$                                                     \\
O: &      $\left(
                            \begin{array}{ccc}
                               -3.45 & -1.28  & -1.40 \\
                               -1.53 & -2.77  & 0    \\
                               -1.61 & -0.07  & -3.34 \\
                            \end{array}
           \right)$    &               $\varepsilon_{0}:$  &
                      $\left(
                            \begin{array}{ccc}
                                 74.61 &    0    &    0    \\
                                   0   &  31.44  &    0    \\
                                   0   &    0    &  70.95  \\
                            \end{array}
                      \right)$
\end{tabular}
\end{ruledtabular}
\end{table}

      The calculated electronic ($\varepsilon_{\infty}$) and static ($\varepsilon_{0}$) dielectric tensors are also given in Table~II. The dielectric tensors are usually overestimated in local density approximation (LDA) calculations due to the underestimation of the band gap. However, application of the scissors correction has previously been found to improve the agreement with experimental data for both rutile and anatase cases.\cite{anatase_phonon} In this study, however, the dielectric tensors are calculated without the scissors correction. Compared with the previous predictions\cite{anatase_phonon,rutile_dielectric} without scissor correction for rutile and anatase phase, we find that the electronic dielectric constant in the $\alpha$-PbO$_{2}$ phase is slightly smaller than those of rutile but larger than the values of the anatase phase, which may be attributed to a more similar structure between $\alpha$-PbO$_{2}$ phase and rutile phase as shown in the Ti$_{2}$O$_{2}$Ti$_{2}$ unit in both phases.

\begin{figure}
\includegraphics[width=4.0cm]{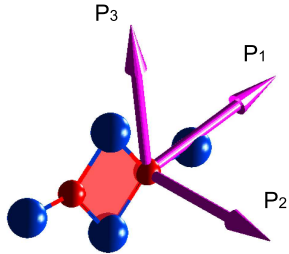}\\
\caption{(Color online) Principal directions of the Born effective charge tensor of the O atom in $\alpha$-PbO$_{2}$-type TiO$_{2}$ (O atoms in red and Ti atoms in blue). The principal values along the three principal directions are $-5.35, -1.19, -3.01$, respectively.}
\label{Fig.3}
\end{figure}

\subsection{Phonons of $\alpha$-PbO$_{2}$-type TiO$_{2}$}

      Since the primitive cell of $\alpha$-PbO$_{2}$-type TiO$_{2}$ contains 12 atoms, there are a total of 36 modes of vibration. The factor group analysis shows that the long wavelength optic phonon modes at the $\Gamma$ point can be decomposed as
\begin{equation*}
\Gamma=4B_{1u}+3B_{2u}+4B_{3u}+4A_{g}+5B_{1g}+4B_{2g}+5B_{3g}+4A_{u}
\end{equation*}

      All the phonon modes are non-degenerate. The $B_{1u}$, $B_{2u}$, $B_{3u}$ modes are IR active, whereas the $A_{g}$, $B_{1g}$, $B_{2g}$ and $B_{3g}$ modes are Raman active and the $A_{u}$ modes are silent. The Raman and IR modes are mutually exclusive due to the presence of inversion symmetry in the crystal.

\begin{table}[htbp]
\caption{ Calculated phonon frequencies ($\omega$) of TiO$_{2}$ at the $\Gamma$ point, and their pressure derivatives ($d\omega/dP$). For the IR-active modes, both TO and LO frequencies are listed.}
\begin{ruledtabular}
\begin{tabular}{ccc|ccc}
IR/Silent & $\omega$ (TO/LO) &  $d\omega/dP$ & Raman & $\omega$ &  $d\omega/dP$  \\
 \hline
         &                  &         & B$_{3g}$ & 154.8 & 0.67 \\
IR       &                  &         & A$_{g}$  & 164.3 & 0.13 \\
$B_{1u}$ &    197.4/229.9   & -2.61   & B$_{1g}$ & 180.1 & 1.71 \\
$B_{3u}$ &    205.9/267.3   &-0.53    & B$_{3g}$ & 215.3 & 0.07 \\
$B_{1u}$ &    253.9/360.1   & 3.62    & B$_{1g}$ & 286.5 & -0.20 \\
$B_{3u}$ &    280.5/286.3   & -0.06   & A$_{g}$  & 295.5 & -0.42 \\
$B_{2u}$ &    301.7/368.0   & 0.54    & B$_{1g}$ & 309.4 & 1.18 \\
$B_{3u}$ &    371.3/477.1   & 5.97    & B$_{2g}$ & 344.6 & 1.00 \\
$B_{2u}$ &    421.3/653.4   & 2.74    & B$_{3g}$ & 388.1 & 2.67 \\
$B_{1u}$ &    427.2/527.8   & 3.14    & B$_{2g}$ & 404.6 & 1.75\\
$B_{3u}$ &    483.1/838.7   & 4.43    & A$_{g}$  & 431.9 & 2.60\\
$B_{1u}$ &    556.5/809.7   & 3.78    & B$_{1g}$ & 445.0 & 4.53\\
$B_{2u}$ &    701.0/788.5   & 3.54    & B$_{3g}$ & 457.5 & 2.60\\
         &                  &         & B$_{3g}$ & 154.8 & 0.67 \\
 Silent  &                  &         & A$_{g}$ & 548.4 & 5.06\\
$A_{u}$  &     303.3        &-0.16    & B$_{1g}$ & 576.3 & 5.35\\
$A_{u}$  &     527.2        &5.00     & B$_{3g}$ & 624.7 & 4.69\\
$A_{u}$  &     630.6        &4.00     & B$_{2g}$ & 813.0 & 3.83
\end{tabular}
\end{ruledtabular}
\end{table}

The calculated vibrational frequencies of the Raman modes, IR TO/LO modes and the silent $A_{u}$ modes, as well as their pressure dependence ($d\omega/dP$) are summarized in Table III. Here $d\omega/dP$ is obtained through a linear fitting of $\omega-P$ data between 0 and 10 GPa, which can be further used to generate the Gr\"{u}neisen parameter for each mode. An average value of 2.29 cm$^{-1}$GPa$^{-1}$ is obtained for $d\omega/dP$ of the Raman modes.

\begin{figure}
\includegraphics[width=9.0cm]{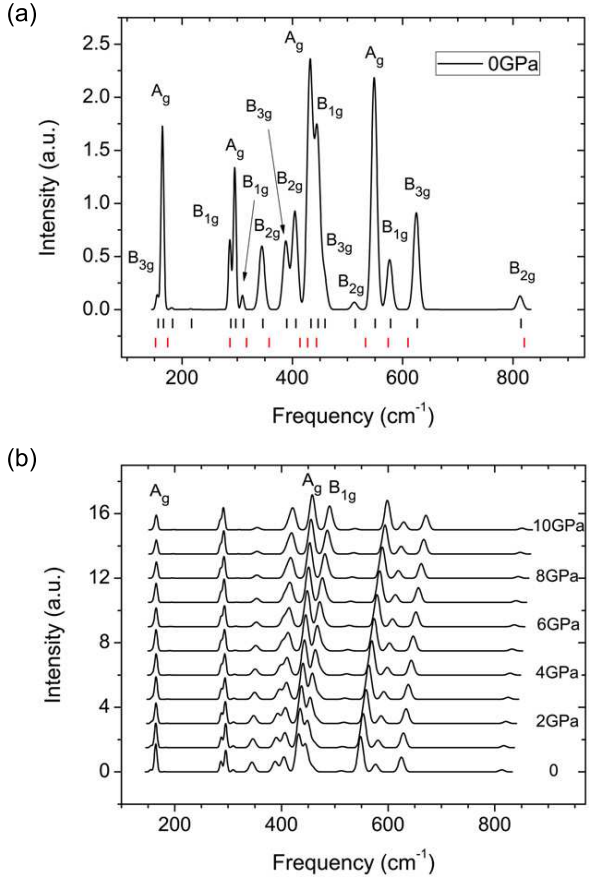}\\
\caption{(Color online) (a) Calculated Raman spectrum of $\alpha$-PbO$_{2}$-type TiO$_{2}$ at 0 GPa. The black and red lines at the bottom of the figure represent the calculated frequencies tabulated in Table III and the experimental results from Ref.~[\onlinecite{TiO2_23}], respectively. (b) Raman spectra at various pressures from 0 to 10 GPa.}
\label{Fig.4}
\end{figure}

To enable a comparison with the experimental results, we calculated Raman spectrum of $\alpha$-PbO$_{2}$-type TiO$_{2}$. The spectrum at 0 GPa is shown in Fig.~4a. The calculated frequencies of Raman active modes in this work and the experimental values by Mammone {\em et al.}\cite{TiO2_23} are indicated by the black and red lines at the bottom of the figure, respectively. By comparing the calculated and the experimental Raman spectra, we conclude that the frequencies at 150 and 172 cm$^{-1}$ observed experimentally\cite{TiO2_23} can be assigned to the $B_{3g}$ (154.8 cm$^{-1}$) and $A_{g}$ (164.3 cm$^{-1}$) modes, respectively. These assignments are further supported by the calculated high pressure Raman spectra shown in Fig.~4b. The gradual overlap of the positions and the decrease in the intensities of the two peaks at high pressures are consistent with the trends observed in the experiment. Particularly, the disappearance of the peak (172 cm$^{-1}$) observed in the experimental Raman spectra was used as an indicator of the occurrence of the phase transition from the $\alpha$-PbO$_{2}$-type TiO$_{2}$ to baddeleyite phase. In addition, the splitting of the two peaks around 450 cm$^{-1}$ ($A_{g}$ and $B_{1g}$) and the significantly stiffening of three peaks above 500 cm$^{-1}$ at high pressures were also clearly recorded in the experiments.\cite{TiO2_23,Phase_Coexistence} It should be noted that, both the calculated and experimental Raman spectra show no soft modes with significant negative derivatives of pressure, as the $B_{1g}$ ($E_{g}$) mode in high pressure Raman spectrum of deformed rutile (anatase) phase. \cite{Strained_rutile,Anatase_TS}

\begin{figure}
\includegraphics[width=9.0cm]{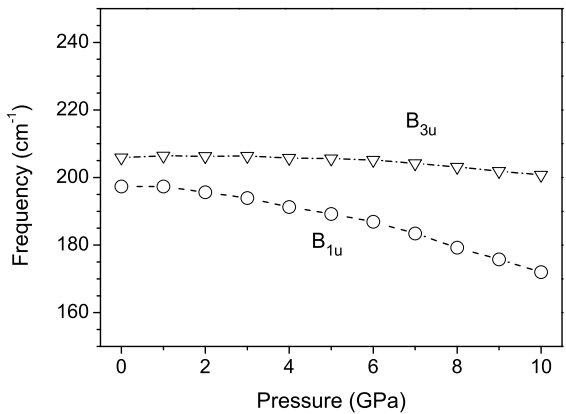}\\
\caption{(Color online) Frequencies of the two IR-active $B_{1u}$ and $B_{3u}$ soft modes as functions of pressure from 0 to 10 GPa.}
\label{Fig.5}
\end{figure}

However, the two lowest IR active modes, $B_{1u}$ at 197.4 cm$^{-1}$ and $B_{3u}$ at 205.9 cm$^{-1}$, respectively, show large negative pressure derivatives as seen in Table III. Although several Raman and silent modes show similar negative-pressure derivatives, those of the two IR modes are much more stronger. These anomalously negative-pressure derivatives suggest a strong softening mechanism which can be correlated to atomic displacements in the structure under pressure and lattice instability. The frequencies of the two IR-active soft TO modes are shown in Fig.~5 as functions of pressure. A nonlinear dependence of the frequencies on pressure is clearly seen.

The eigenvectors of the two soft modes $B_{1u}$ and $B_{3u}$ are shown in Fig. 6. The vibration patterns can be analyzed based on the Ti$_{2}$O$_{2}$Ti$_{2}$ complex consisting of two Ti$_{3}$O complexes through sharing a Ti$_{2}$ edge as shown in the Fig. 1. The atomic vibrating patterns of the Ti$_{3}$O(Ti$_{3}$O-I and Ti$_{3}$O-II) and Ti$_{2}$O$_{2}$Ti$_{2}$ complexes are shown in the left part of the figure. The vertical movement of the oxygen atom out of the plane formed by the titanium atoms can be clearly observed. On the other hand, the oddness character(denoted by ``$u$" subscript) of $B_{1u}$ and $B_{3u}$ under inversion requires that those atoms connecting with each other through the inversion symmetry operation have the same vibration components. Thus the two edge Ti atoms, connected with inversion operation, vibrate in phase but slightly deviate with the vibrating direction of the two end Ti atoms in the Ti$_{2}$O$_{2}$Ti$_{2}$ unit. Similarly, the two oxygen atoms locating on the two sides of the Ti$_{4}$ plane also have the same vibrating component and they vibrate vertically to the Ti$_{3}$ plane.

\begin{figure}
\includegraphics[width=8.0cm]{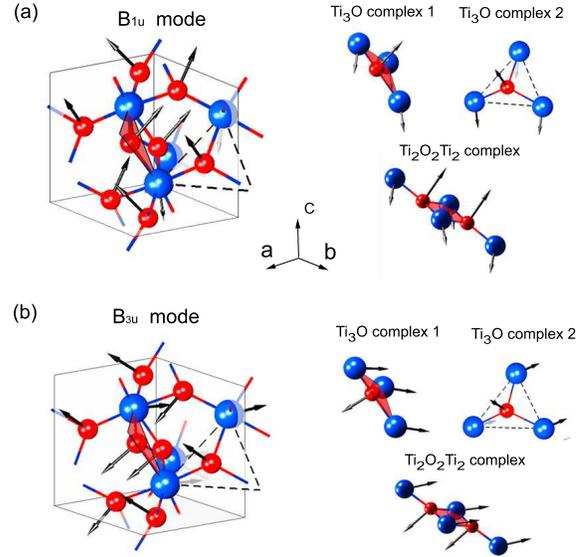}\\
\caption{(Color online) Eigenvectors of the $B_{1u}$ (a) and $B_{3u}$ (b) soft modes. The vibrational eigenvectors of the atoms in the Ti$_{3}$O and Ti$_{2}$O$_{2}$Ti$_{2}$ complexes (O atoms in red and Ti atoms in blue) are given on the right side of each figure. }
\label{Fig.6}
\end{figure}

      To find out whether there exists any other soft phonon modes, we calculated the phonon dispersion curves and the vibration density of states (DOS) of the $\alpha$-PbO$_{2}$-type phase at 0 GPa and 10 GPa, respectively. As shown in Fig.~7a, a significant shift of the spectrum towards the high frequency side is seen at frequencies above 300 cm$^{-1}$ which is consistent with the larger pressure dependence of the high frequency phonon modes given in Table~III. Fig.~7b shows details in the region between 150 and 250 cm$^{-1}$. These branches are related to the soft $B_{1u}$ and $B_{3u}$ modes. It is clear that in addition to the $\Gamma$ point, frequencies of the modes along the $\Gamma$-$Y$,$\Gamma$-$X$,$\Gamma$-$Z$ and parts of $Y$-$U$, $Y$-$S$, $X$-$S$, $Z$-$T$ also decrease at high pressure.

\begin{figure}
\includegraphics[width=9.0cm]{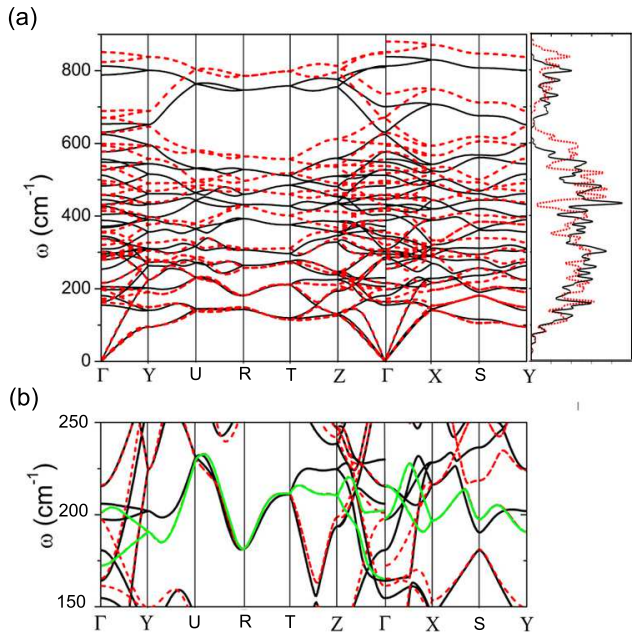}\\
\caption{(Color online) (a) Calculated phonon dispersion curves of $\alpha$-PbO$_{2}$-type TiO$_{2}$ at 0 GPa (black solid lines) and 10 GPa (red dashed lines). The phonon density of states (DOS) at the two pressure values are plotted in the left panel. (b) Details of the region from 150 to 250 cm$^{-1}$ in (a). The vibrational branches associated with the IR active $B_{1u}$ and $B_{3u}$ modes at 10 GPa are drawn in green lines to show mode softening. }
\label{Fig.7}
\end{figure}

\subsection{Structure and Phonons of baddeleyite TiO$_{2}$}

\begin{figure}
\includegraphics[width=7.0 cm]{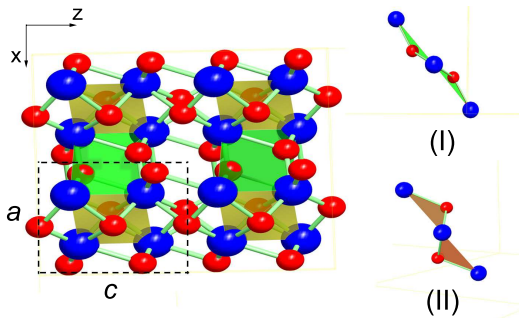}\\    
\caption{(Color online) Crystal structure of baddeleyite-type TiO$_{2}$ (O atoms in red and Ti atoms in blue). The two types of Ti$_{2}$O$_{2}$Ti$_{2}$ complexes in the structure are shown in brown color (type I) and green color (type II), respectively. The dashed lines depict the unit cell of the structure. The right panel shows the side views of Ti$_{2}$O$_{2}$Ti$_{2}$-I and Ti$_{2}$O$_{2}$Ti$_{2}$-II.}
\label{Fig.8}
\end{figure}

The atomic structure of the baddeleyite TiO$_{2}$ (monoclinic) is illustrated in Fig.~8. The CN of Ti atom in this structure is 7, compared to 6 in $\alpha$-PbO$_{2}$-type TiO$_{2}$. All Ti atoms are equivalent and occupy the $4e$ Wyckoff sites, but the 8 oxygen atoms can be divided into two inequivalent groups, OI and OII, which occupy two sets of $4e$ Wyckoff sites respectively. A structure similar to the Ti$_{2}$O$_{2}$Ti$_{2}$ complex in $\alpha$-PbO$_{2}$ phase also exists in baddeleyite TiO$_{2}$, as shown in Fig.~8. However, in baddeleyite TiO$_{2}$, there are two types of Ti$_{2}$O$_{2}$Ti$_{2}$ complexes, termed as Ti$_{2}$O$_{2}$Ti$_{2}$-I and Ti$_{2}$O$_{2}$Ti$_{2}$-II, respectively, depending on the type of oxygen (OI or OII) involved. The distances from the oxygen atoms to the Ti$_{4}$ plane in the two types of Ti$_{2}$O$_{2}$Ti$_{2}$ complexes are different. Using the experimental lattice parameters of $a=4.640$ {\AA}, $b=4.760$ {\AA}, and $c=4.810$ {\AA},\cite{baddeleyite_structure} the OI and OII atoms in the relaxed structures are 0.358 {\AA} and 0.940 {\AA} from the Ti$_4$ plane, both being much larger than the corresponding value of 0.12 {\AA} in the $\alpha$-PbO$_{2}$-type TiO$_{2}$ at zero pressure.

Finally, the calculated phonon frequencies at the $\Gamma$ point of baddeleyite TiO$_{2}$ are listed in Table IV. Fifteen IR active (8$A_{u}$+7$B_{u}$) and eighteen Raman active (9$A_{g}$+9$B_{g}$) modes are expected for baddeleyite phase from the factor group analysis.

\begin{table}[htbp]
\caption{\label{tab:results} Calculated phonon frequencies of IR active modes ($A_u$ and $B_u$) and Raman active modes ($A_g$ and $B_g$) for baddeleyite TiO$_{2}$ at the $\Gamma$ point. For the IR-active modes, both TO and LO frequencies are listed.}
\begin{ruledtabular}
\begin{tabular}{ll}
IR/Raman   &   Frequency(cm$^{-1}$)                             \\     %
\hline
$A_u$ (TO/LO)    &  227.9/242.8, 246.1/256.7, 331.9/332.4,      \\
                 &   384.1/404.5, 439.7/540.5, 451.6/452.3,     \\
                 &  582.9/860.7, 733.3/736.3                    \\
$B_u$ (TO/LO)    &  110.5/196.3, 295.8/341.1, 359.2/457.9,      \\
                 & 495.7/520.0, 529.9/560.8, 561.5/702.5,       \\
                 &  784.7/960.0                                 \\
$A_g$            &  136.8, 217.0, 249.7, 348.7, 399.4, 427.0,   \\
                 & 511.7, 647.3, 678.5                          \\
$B_g$            &  189.1, 296.6, 352.8, 423.2, 430.4, 493.5,   \\
                 & 621.7, 707.0, 814.9
\end{tabular}
\end{ruledtabular}
\end{table}

\section{DISCUSSION}
The transition from the $\alpha$-PbO$_{2}$-type phase to the baddeleyite phase mediated through the soft phonon modes can be derived directly from the comparison of the related softening modes and the
structural features of baddeleyite phase. The significant deviation of the oxygen atoms from the Ti$_{4}$ plane in the Ti$_{2}$O$_{2}$Ti$_{2}$ complex of the baddeleyite phase arises from relaxation of oxygen atoms along the directions described by the eigenvectors of the $B_{1u}$ and $B_{3u}$ soft modes in $\alpha$-PbO$_{2}$ TiO$_2$. This soft phonon mode based mechanism was further confirmed by results of our molecular dynamics (MD) simulation of the $\alpha$-PbO$_{2}$-type TiO$_{2}$ which was carried out at experimental conditions of 300 K and 37 GPa.\cite{TiO2_23}
The MD simulation was implemented by the ``General Utility Lattice Program"(GULP) package\cite{GULP} with a NPT ensemble using Dreiding empirical force potential\cite{Dreiding} which has previously been adopted in simulations of molecular adsorption on TiO$_2$ surface,\cite{MD_TiO2} zeolite confined Ti(OH)$_{4}$ nanoparticles\cite{Zeolite} and other surfactant interactions at mineral surfaces.\cite{mineral_1} The Nos\'{e}-Hoover thermostat was applied to control the temperature. It should be noted that the initial velocities, which were produced randomly with probabilities based on Gaussian distribution, are not related to the mode direction of $B_{1u}$ and $B_{3u}$ soft modes. The initial structure and the snapshot of a representative intermediate structure of $\alpha$-PbO$_{2}$-type TiO$_{2}$ are shown in Fig.~9.\cite{note} Strong distortions in the lattice and out-of-plane displacement of the oxygen atoms can be observed. The single type of Ti$_{2}$O$_{2}$Ti$_{2}$ complex in the $\alpha$-PbO$_{2}$ phase evolves into two types (I and II) which are similar to those in the baddeleyite structure given in Fig.~8. The relaxation of the Ti$_{2}$O$_{2}$Ti$_{2}$ complex which shows characteristic features of the $B_{1u}$ or $B_{3u}$ soft phonon induces significant shear deformation in the $y$-$z$ plane. It should be pointed out that a complete structural transformation from the $\alpha$-PbO$_{2}$ phase to the baddeleyite phase was not observed in this simulation. This is due to constraints of the supercell ($2\times 2\times 2$) and possible high potential barrier along the pathway. Nevertheless, result of the molecular dynamics simulation does indicate that the local atomic displacements are closely related to the $B_{1u}$ and $B_{3u}$ soft modes at the $\Gamma$ point.

\begin{figure}
\includegraphics[width=8.0cm]{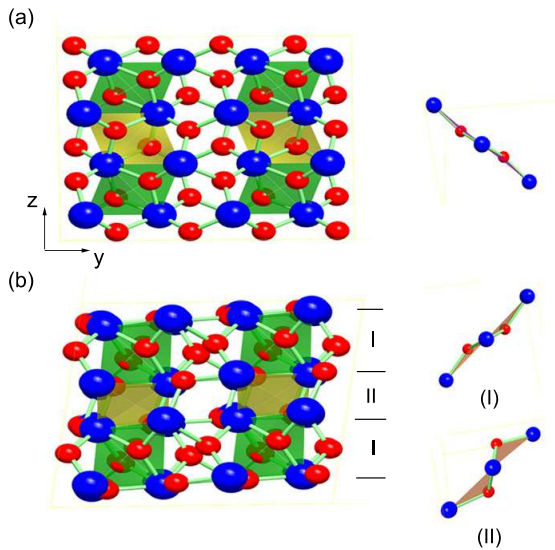}\\
\caption{(Color online) Comparison of the structure of $\alpha$-PbO$_{2}$-type TiO$_{2}$(a) with that of the deformed $\alpha$-PbO$_{2}$-type TiO$_{2}$(b) obtained from molecular dynamics simulations. The right panel shows the side views of the structures of Ti$_{2}$O$_{2}$Ti$_{2}$. Red(small) balls and blue(big) balls represent oxygen and titanium atoms, respectively.}\label{Fig.9}
\end{figure}

In this study, the structural difference between the $\alpha$-PbO$_{2}$-type and the baddeleyite phases is analyzed based on the basic building units Ti$_{3}$O and Ti$_{2}$O$_{2}$Ti$_{2}$ complexes, in contrast to the commonly used TiO$_{6}$ octahedra.\cite{Octahedra_1,Octahedra_2} One clear advantage of using Ti$_{3}$O and Ti$_{2}$O$_{2}$Ti$_{2}$ as the basic units is in the description of the $B_{1u}$ and $B_{3u}$ soft modes of the $\alpha$-PbO$_{2}$-type phase, and the correlation between the phase transition and the out-of-plane displacements of the oxygen atoms in these soft modes. As a matter of fact, such soft modes not only dictate phase transition initiated from the $\alpha$-PbO$_{2}$-type phase, similar features have  also been observed in the high pressure phase transitions originating from rutile and anatase phases of TiO$_{2}$. The $B_{1g}$ soft mode in rutile phase and the $E_{g}$ soft mode in anatase phase can be described similarly in terms of Ti$_{3}$O and Ti$_{2}$O$_{2}$Ti$_{2}$ complexes, where out-of-plane displacements of the oxygen atoms relative to the Ti$_{3}$ plane are observed.\cite{Rutile_Cacl2,Raman_Anatase} For instance, in the case of rutile to $\alpha$-PbO$_{2}$ phase transition, the displacements of the O atoms associated with the $B_{1g}$ mode lead to the CaCl$_{2}$ structure with corrugated Ti$_{2}$O$_{2}$Ti$_{2}$ unit chains along the [110] direction, and the final $\alpha$-PbO$_{2}$-type structure can be achieved by a subsequently shearing in the (011) plane of the CaCl$_{2}$-type structure.\cite{Rutile_Cacl2,Pathways,alpha_PbO2_TS1,alpha_PbO2_TS2} In the anatase case, the high pressure transition to the $\alpha$-PbO$_{2}$ phase has also been found to be closely related to the soft $E_{g}$ mode, and the gliding of anionic layers results in the transformation.\cite{Anatase_TS} Thus, the instability of the Ti$_{3}$O units may be regarded as a common feature of high pressure transformations of TiO$_{2}$.

\section{CONCLUSIONS}
Softening of phonon modes, which is usually an important indicator of structural instability during displacive phase transition, has previously been found in both rutile and anatase TiO$_{2}$ under high pressure. However, for phase transitions originating from the $\alpha$-PbO$_{2}$-type TiO$_{2}$, no soft modes have yet been reported during high pressure transition. Using the state-of-art DFPT calculations, we show that there exist two IR-active soft phonon modes ($B_{1u}$ and $B_{3u}$) around 200 cm$^{-1}$ which are absent in the high pressure Raman spectrum. These two modes play an important role in understanding the atomic displacements in  $\alpha$-PbO$_{2}$-type TiO$_{2}$ and transformation to baddeleyite polymorph. In addition to zone-center frequencies, the changes in phonon dispersions with pressure indicate that significant softening also occurs in the zone-boundary modes, which suggests that gliding of layers in $\alpha$-PbO$_{2}$-type TiO$_{2}$ is likely to occur during the $\alpha$-PbO$_{2}$-type TiO$_{2}$ phase transition.
Furthermore, the symmetry assignments of the Raman peaks for $\alpha$-PbO$_{2}$-type TiO$_{2}$ are given for the first time. Since there is usually a coexistence of different phases during the process of high pressure phase transitions, the calculated Raman spectrum and its change with pressure are valuable for experimentalist to identify the onset of phase transition related to the $\alpha$-PbO$_{2}$-type phase.

Acknowledgement: Computations were performed at the Centre for Computational Science and Engineering(CCSE) and High Performance Computing Centre at NUS.

\end{document}